\documentclass[utf8]{frontiersinFPHY_FAMS_1}
\usepackage{url,hyperref,lineno,microtype,subcaption}
\usepackage[onehalfspacing]{setspace}

\def\keyFont{\fontsize{8}{11}\helveticabold }
\def\firstAuthorLast{Kulchukova {et~al.}} 
\def\Authors{Zhamila Kulchukova\,$^{1,2,3}$, Albert Ferrando\,$^{4}$ and Anton S. Desyatnikov\,$^{3,*}$}
\def\Address{$^{1}$Fundamentale Physik für Metrologie, Physikalisch-Technische Bundesanstalt, Braunschweig, Germany \\$^{2}$Institute of Mathematical Physics, Technische Universität Braunschweig, Braunschweig, Germany\\$^{3}$Department of Physics, School of Sciences and Humanities, Nazarbayev University, Astana, Kazakhstan \\ $^{4}$Instituto de Ciencia de los Materiales, Universidad de Valencia, Paterna, Spain }
\def\corrAuthor{Anton S. Desyatnikov, Department of Physics, School of Sciences and Humanities, Nazarbayev University, 53 Kabanbay Batyr Av., Astana 01000, Kazakhstan}
\def\corrEmail{anton.desyatnikov@nu.edu.kz}

\usepackage{etoolbox}

\makeatletter
\AtBeginDocument{%
  \protected\def\doi#1{\@ifnextchar.{\@gobble}{}}
}
\makeatother

\begin{document}
\onecolumn
\firstpage{1}

\title[Vortex rings]{Catastrophes in Reconnections of Optical Vortex Rings} 

\author[\firstAuthorLast ]{\Authors} 
\address{} 
\correspondance{} 

\extraAuth{}

\maketitle

\begin{abstract}
A small change in parameters leads to a small change in the outcomes, in most cases. For instance, an optical vortex ring, created in the superposition of two laser beams, can slightly shrink if the phase changes minutely. However, dynamical systems can also experience significant changes due to small perturbations in parameters; the vortex rings can reconnect, nucleate or collapse into a point. Parameters that cause such striking outcomes are at the focal point of catastrophe theory - a branch of bifurcation theory that studies degenerate critical points of a potential function. We present theoretical study of topological configurations of optical vortex rings in elliptic Gaussian beams, which change dramatically at specific beam parameters. We identify two types of catastrophes, cusp and butterfly, near which very small changes of elliptic deformations lead to complex topological transformations of vortex rings. 

{\bf Methods:} Analytical derivations and numerical simulations for bifurcation diagrams and topological transformations.

\tiny
 \keyFont{ \section{Keywords:} structured light, optical vortex rings, vortex reconnections, Gaussian beams, catastrophe theory} 
\end{abstract}

\bigskip
\section{Introduction}
\label{intro}

Structured light, i.e. the light field tailored into a specific structure in its polarization, spatial, temporal and spectral domains - has been gaining significant interest in recent decade~\cite{Rubinsztein-DunlopEtal2017, ForbesOlivieraDennis2021, PetersCocotosForbes2025}. The primary example of spatially structured light are Laguerre-Gaussian and Bessel beams carrying optical vortices. The optical vortex, thread of darkness, is a line in three dimensional light fields at which the intensity vanishes and phase becomes singular. Originally named ``phase dislocation'' due to analogy with dislocations in crystalline materials~\cite{NyeBerry1974, Wolf2001}, the term ``optical vortex'' was later coined due to resemblance to fluid vortices~\cite{CoulletGilRocca2010,BazhenovVasnetsovSoskin1990}. Light surrounding the vortex core twists and rotates, carrying well-defined orbital angular momentum (OAM), which is utilized across a wide range of applications, from particle trapping to nonlinear optics, optical communications, quantum cryptography, imaging, spectroscopy, metrology, and more~\cite{PiO2005,DennisEtal2009,SoskinEtal2017,ShenEtal2019}.

In Laguerre-Gaussian continuous-wave beams the OAM is longitudinal and intrinsic, i.e. the OAM vector coincides with the mean momentum and is invariant to the choice of coordinates. Similar structure is realised with short optical pulses in dispersive media~\cite{PRE2000, MalomedPRE2000, Malomed2005,EilenbergerEtal2013}. However, that is not the only possible configuration. In nonlinear media, collapse arrest and filamentation of ultrashort pulses generates spatiotemporal optical vortices (STOVs) and {\em vortex rings}~\cite{JhajjEtal2016,LeEtal2024}, which carry transverse OAM~\cite{HancockZahedpourMilchberg2021}. The nature of transverse OAM in STOVs has recently sparked a debate about its origins and interpretation~\cite{Bliokh2021,Porras2024, HancockEtal2024,Tripathi2025, VoBliokhAlonso2026}. The discovery of STOVs has introduced a new class of structured light in which space and time are coupled to generate complex light topologies~\cite{Wan2022,ShenEtal2023, Bekshaev2024,LiuCaoZhan2024}. This has opened potential applications in light–matter interactions and intense-light phenomena, particularly with the emergence of techniques for reconstructing experimentally generated STOVs~\cite{ChenEtal2021}.

Although STOV rings where observed to appear spontaneously in nonlinear media~\cite{JhajjEtal2016,LeEtal2024}, similar effect were predicted for continuous-wave beams~\cite{Knots2012}, and, in fact, can be traced back to such fundamental optical phenomenon as Airy rings~\cite{BoivinWolf1965, BoivinDowWolf1967}. It is not surprising that the topological transformations of stationary vortex rings, such as nucleation, annihilation, and reconnections~\cite{BerryDennis2001b, BerryDennis2001a, BerryDennis2007}, were also observed with STOV rings, creating complex connected OAM states~\cite{LeEtal2024,Adams2024,AdamsChong2025}. The universality of these phenomena stimulates a search for a basic and simple model, such as destructive interference of a Gaussian beam with a plane wave~\cite{Desyatnikov2023, KulchukovaDesyatnikov2024}. This model captures the main topological effects of optical vortex rings while remaining experimentally accessible~\cite{FerrandoEtal2023}. Despite its simplicity, this system manifests a rich variety of vortex rings transformations and unexpected complexity of topological phase diagram. In particular, in our recent studies~\cite{Zham25,Zham26}, we observed that the bifurcation lines in the parameter space can fold into complex patterns reminiscent of Thom’s elementary catastrophes~\cite{Thom1972}.

Catastrophe theory is not new to optics, or physics in general. At its core, catastrophe theory considers how dynamical systems respond to small changes in parameters. However, in some systems, perturbing parameters can cause dramatic changes - such events are called catastrophes. Catastrophe theory was first introduced by Thom~\cite{Thom1972} and later developed thanks to the efforts of Arnold~\cite{ArnoldWassermann2003} and Zeeman~\cite{Zeeman1979}. In the presence of catastrophes, the critical points of a potential function, that describes the dynamics of the system, can appear or disappear as the parameters are varied. Critical points at which the first and one or more higher-order derivatives vanish are called degenerate critical points, or singularities, of the potential function. Mapping singularities onto the space of control parameters form bifurcation lines and reveals one of the well-defined patterns - the seven elementary catastrophes, as classified by Thom: fold, cusp, swallowtail, butterfly, hyperbolic umbilic, elliptic umbilic, and parabolic umbilic.

Thom's catastrophes are not limited to parameter spaces either. Optical caustics are direct manifestations of catastrophes in real space, which are envelopes of divergent intensity formed by light reflecting or refracting off a curved surface, much like the speckles of light at the bottom of the pool. In some sense, they are the optical equivalent of rogue waves that form on the open water, and can provide a solid framework for studying extreme wave phenomena~\cite{HöhmannEtal2010, MathisEtal2015}. Caustics are stable under perturbations, unlike focused light, which may get destroyed if the optical apparatus is deformed~\cite{Nye1999, BerryUpstill1980}, and can also be integrated into structured light beams~\cite{ZannottiEtal2017, ZannottiDiebelDenz2017,ZannottiEtal2020, TengEtal2021}. The shape of the caustics can be modeled using geometrical optics principles; however, in the microscopic scale around the brightest point, complex diffraction patterns are revealed~\cite{TrinkausDrepper1977, Pearcey1946, Berry1976, Borghi2011, Borghi2016, Borghi2022}. Diffraction patterns of caustics depicting one of the seven elementary catastrophes have been extensively studied, theoretically and experimentally~\cite{Nye1978, BerryNyeWright1978, Nye1986, Nye1998, Nye2003, Nye2007, Nye2008}. Analyzing such patterns, one can notice an arrangement of dark points with zero intensity - optical vortices - that acts as a skeleton supporting the caustic, as if to balance the extreme intensity of light~\cite{Berry1976}. 

In this paper, we attempt a systematic approach to reveal and explore catastrophes in the topological phase diagrams of optical vortex rings generated in the superpositions of elliptic and astigmatic Gaussian beams with a co-propagating plane wave. Catastrophes appear as folding of bifurcation lines of topological events when the ellipticity and astigmatism parameters of the beam reach degenerate critical points of corresponding phase functions. In the next section we introduce our main framework, followed by the section devoted to the detailed studies of the appearance of catastrophes as well as their manifestations in topological transformation sequences, driven by the change of parameters. Finally, we combine our findings in a single diagram presenting a map of catastrophes in parameter space.  

\section{Vortex rings and reconnections}

Our basic model, which leads to the nucleation and reconnections of optical vortex rings, is based on the destructive interference of a Gaussian beam and a plane wave~\cite{Desyatnikov2023,KulchukovaDesyatnikov2024}. Propagation in free space of a laser beam with the slowly-varying envelope $E(x,y,t)$ is governed by the Schr\"odinger-type paraxial wave equation, $4i\partial_tE+\partial^2_{x}E+\partial^2_{y}E=0$, here the dimensionless transverse spatial coordinates $x$ and $y$ are normalized by the arbitrary scale $w_0$ and the evolution variable $t=z/z_d$ is the propagation distance $z$ measured in diffraction lengths $z_d=kw_0^2/2$ with the wavenumber $k=2\pi/\lambda$ and the carrier wavelength $\lambda$. Optical vortices are found as the field nodal lines $E=0$ of the superposition, $E(x,y,t)=G(x,y,t)-Pe^{i\phi}$, with constant amplitude $P$ and phase $\phi$ for the plane wave. The Gaussian beam is assumed to have an elliptic cross-section~\cite{KulchukovaDesyatnikov2024},
\begin{equation}
    G(x,y,t)=\sqrt{\frac{w_{0x}w_{0y}}{w_x(t)w_y(t)}}\exp\left(-x^2\left(\frac{1}{w_x^2(t)}-\frac{i}{R_x(t)}\right)-y^2\left(\frac{1}{w_y^2(t)}-\frac{i}{R_y(t)}\right)-i\chi(t)\right),
\end{equation} 
here the geometric Gouy phase $\chi(t)$,
\begin{equation}
    \chi(t)=\frac{1}{2}\left(\tan^{-1}\frac{t-d/2}{2\cos^2\epsilon}+\tan^{-1}\frac{t+d/2}{2\sin^2\epsilon}\right).
\end{equation}
The beam widths in the transverse plane $w_{x,y}(t)$ and the wavefront radii $R_{x,y}(t)$ are given by:
\begin{align*}
&\frac{1}{w_x^2(t)}=\frac{2\cos^2\epsilon}{4\cos^4\epsilon+(t-d/2)^2}, &\frac{1}{w_y^2(t)}=\frac{2\sin^2\epsilon}{4\sin^4\epsilon+(t+d/2)^2},\\
&\frac{1}{R_x(t)}=\frac{t-d/2}{4\cos^4\epsilon+(t-d/2)^2}, &\frac{1}{R_y(t)}=\frac{t+d/2}{4\sin^4\epsilon+(t+d/2)^2}.
\end{align*}
There are two key parameters, which determine the beam profile: (i) the \textit{ellipticity} $\epsilon\in(0,\pi/2)$ controls the ratio of the beam waists $w_{0x}=\sqrt{2}\cos\epsilon$ and $w_{0y}=\sqrt{2}\sin\epsilon$, and (ii) the distance between the focal planes, $d\in\mathbb{R}$, serves as \textit{astigmatism} parameter; the focal $x$ and $y$ planes are located at $t=\pm d/2$, respectively.

Superposition of such beams with a plane wave creates vortex rings that appear deformed in three dimensions, except when they are flat in the special transverse plane located at $t_*=-d/(2\cos2\epsilon)$. In our previous work~\cite{KulchukovaDesyatnikov2024}, we identified three distinct topological transformations of vortex rings in elliptic Gaussian beams: (i) on-axis nucleation of a single ring, (ii) off-axis nucleation of a pair of rings, and (iii) rings reconnections. The latter two topological transformations are possible due to the elliptic deformation of the beam by $\epsilon$ and $d$, in contrast to the radially symmetric beam recovered for $\epsilon=\pi/4$ and $d=0$~\cite{Desyatnikov2023}.

Starting with the amplitude equation, $|G(x,y,t)|=P$, rearranging the terms results in the conic equation describing the amplitude iso-surfaces:
\begin{equation}
    \frac{x^2}{w_x^2(t)}+\frac{y^2}{w_y^2(t)} = F(t), \label{eq:conicG}
\end{equation}
here the function $F(t)$ contains the parameter $P$,
\begin{equation}\label{eq:F}
    F(t) = -\ln \left(P\sqrt{w_x(t)w_y(t)/\sin 2\epsilon}\right).
\end{equation}
In a similar manner, we reformulate the phase equation, $\arg G(x,y,t)=\phi$,
\begin{equation}
    \frac{x^2}{R_x(t)}+\frac{y^2}{R_y(t)} = \phi + \chi(t)+2\pi n,\;n\in\mathbb{Z}. \label{eq:conicH}
\end{equation}

The intersection of two conics described by Eqs.~\eqref{eq:conicG} and \eqref{eq:conicH} in each transverse plane located at $t$ corresponds to the field nodes (optical vortices) at four points $(\pm x,\pm y)$, due to quadrupole symmetry~\cite{FerrandoEtal2023}. It follows that topological events will occur in either the $x=0$ or $y=0$ plane. Consequently, the topological events in $x=0$ plane are independent of the events at $y=0$; the only instance when they intersect is at the optical axis. Therefore, we consider these events separately by combining the $x$ and $y$ terms of Eqs.~\eqref{eq:conicG} and \eqref{eq:conicH}:
\begin{equation}
    \begin{split}
        \phi+2\pi n &= f_x(t) \equiv \frac{t-d/2}{2\cos^2\epsilon} F(t) - \chi(t),\\
        \phi+2\pi n &= f_y(t) \equiv \frac{t+d/2}{2\sin^2\epsilon} F(t) - \chi(t), \label{eq:f1f2defined}
    \end{split}
\end{equation}
thus $f_x(t)$ and $f_y(t)$ describe phase relations at the surface Eq.~\eqref{eq:conicG} in $y=0$ and $x=0$ planes, respectively.

It follows that, at the optical axis $x=y=0$, nucleations of vortex rings occur at the on-axis locations $t$ solving the system
\begin{equation} \label{eq:onaxis}
    \begin{cases}
        P=\sqrt{\sin 2\epsilon/w_x(t)w_y(t)}\\
        \phi+2\pi n=-\chi(t).
    \end{cases}
\end{equation}

\section{Catastrophes}

We aim here to explore the self-folding and increased complexity of topological domains with different number of vortex rings, observed with increasing $\epsilon$ or $d$ beyond the degree of ``moderate deformation''~\cite{KulchukovaDesyatnikov2024}. This degree, qualitatively, can be defined as follows. The amplitude iso-surfaces of the fundamental Gaussian mode can be categorized into two shapes: a simple spheroid with one maximum at $t=0$ and a ``dumbbell'' with one minimum at $t=0$ and two maxima~\cite{Desyatnikov2023}. The elliptical Gaussian beam is of ``moderate deformation'' if the amplitude iso-surfaces of the beam resemble the surfaces of the radially symmetric beam. As we discuss below, departure from this definition leads to novel topological reactions.

To understand these transitions, we employ principles of catastrophe theory and analyze the characteristic phase functions in the whole parameter space $(\epsilon,d)$. Analyzing zeros of the first and subsequent derivatives of the potential functions can reveal degenerate critical points in the parameter space, around which the system experiences dramatic changes. 

Reactions on the optical axis Eq.~\eqref{eq:onaxis} occur at $F(t)=0$ in Eq.~\eqref{eq:F}. The first and second derivatives of $F(t)$ are polynomials of degree 3 and 6, respectively, and the parameter $P$ disappears in both. Solving for zeros of the first two derivatives determines the bifurcation lines in the parameter space $(\epsilon,d)$ that change the number of roots of $F(t)=0$, thus changing the number of vortices,
\begin{equation} \label{eq:cuspeqs}
    \begin{cases}
        \partial_tF(t)=0 \\
        \partial_t^2F(t)=0.
    \end{cases}
\end{equation}

Next, the critical points of $f_x(t)$ and $f_y(t)$ in Eq.~\eqref{eq:f1f2defined} indicate the parameters at which topological reactions occur off-axis. The first derivatives of these functions retain dependence on $P$ and, therefore, we consider higher orders, because the bifurcations occur as double zeros of the lowest parameter-free derivative. The second and third derivatives turn the equations of characteristic phases into polynomials of order 5 and 8, respectively, and also eliminate both parameters $P$ and $\phi$:
\begin{align} \label{eq:butterflyeqs}
  \begin{cases}
    \partial_t^2 f_x(t)=0 \\
    \partial_t^3 f_x(t)=0,
  \end{cases}
  \quad
  \begin{cases}
    \partial_t^2 f_y(t)=0 \\
    \partial_t^3 f_y(t)=0.
  \end{cases}
\end{align}
The solutions are the double roots of the quintic numerator of the first equation; as we discuss below, they change the number of cusp catastrophes in the $(P,\phi)$ plane thus determine bifurcation lines in the parameter space $(\epsilon,d)$.

We solve the systems of Eqs.~\eqref{eq:cuspeqs} and~\eqref{eq:butterflyeqs} numerically and verify predictions of catastrophe theory with specific parameters of the Gaussian beam and then build the full picture in the parameter space $(\epsilon,d)$.

\subsection{Elliptic beams}

In this subsection we focus on purely elliptic beams with $d=0$. The parameter $\epsilon$, as mentioned previously, controls the ratio of waists of the beam on the $x=0$ and $y=0$ planes; the symmetries of such beams were discussed in greater detail previously~\cite{KulchukovaDesyatnikov2024}. To reiterate, at the focal plane $t=0$, the elliptic transverse profile of the beam is elongated along $y=0$ if $\epsilon<\pi/4$ and along $x=0$ if $\epsilon>\pi/4$. At $t_c=\pm\sin2\epsilon$ the beam has a circular cross-section and the ellipticity inverts at $|t|>|t_c|$.

\begin{figure*} 
    \centering
    \includegraphics[width=0.99\linewidth]{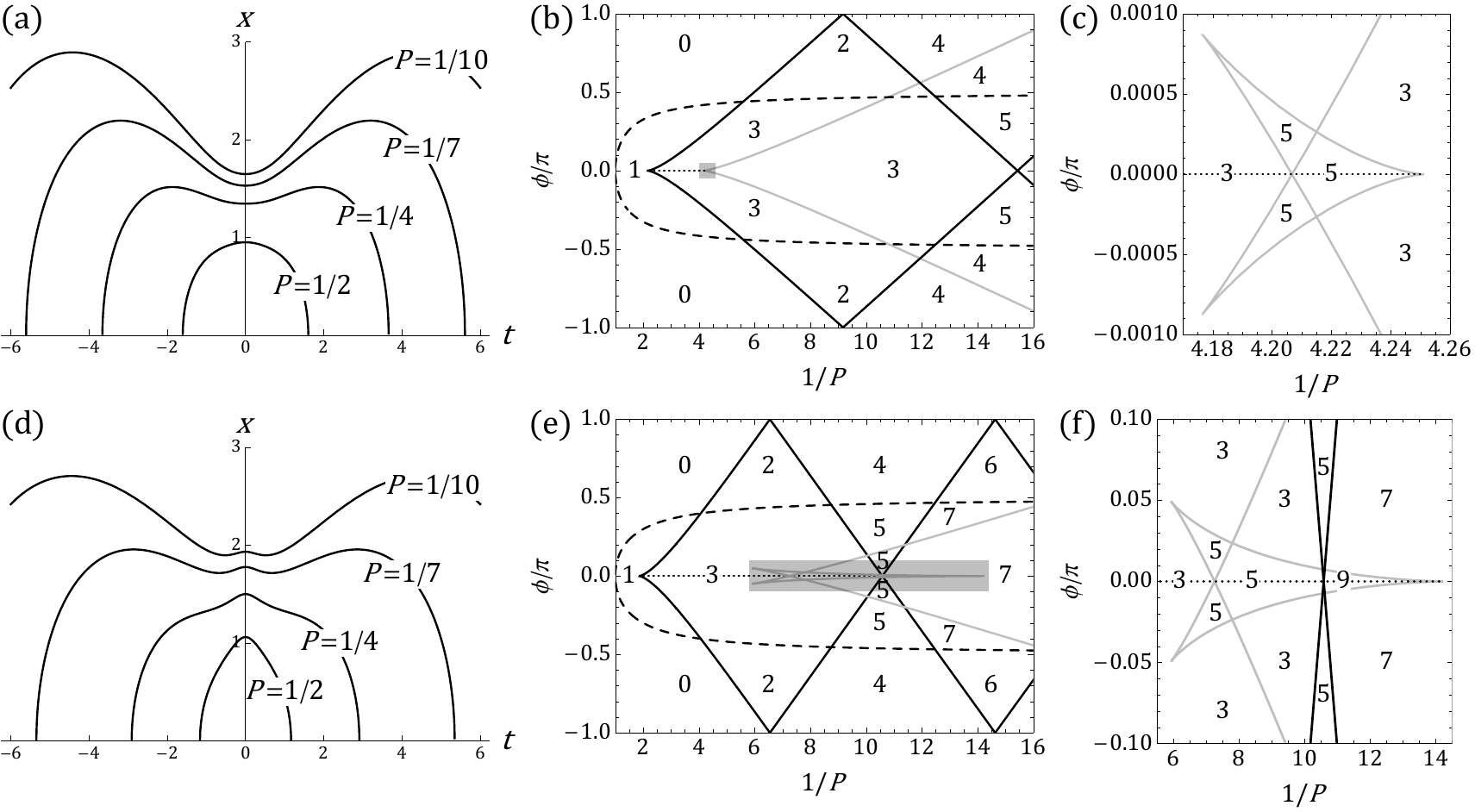}
    \caption{(a,d) Contours of amplitude iso-surfaces of the elliptic Gaussian beams ($d=0$); ellipticity $\epsilon=\pi/5$ for the top row (a-c) and $\epsilon=\pi/7$ for the bottom row (d-f). (b,e) The number of vortex rings in the parameter domains $(P,\phi)$ separated by topological events located at the optical axis (dashed bifurcation lines), at the $x=0$ plane (black solid lines), the $y=0$ plane (grey solid lines), and at the focal plane (dotted lines). (c,f) Magnified regions of the corresponding grey parameter domains in (b,e). }
    \label{fig:cats in elliptic}
\end{figure*}

Solving on-axis Eq.~\eqref{eq:cuspeqs} with $d=0$ we did not identify new bifurcations, however, the off-axis systems Eq.~\eqref{eq:butterflyeqs} predict the appearance of critical points of $f_x(t)$ in the $y=0$ plane for $\epsilon < \pi/4.78$ and critical points of $f_y(t)$ in the $x=0$ plane for $\epsilon>\pi/3.44$, and, therefore, we should observe dramatic changes to the system near these specific values. In the following we will focus on bifurcations in the $y=0$ plane, Fig.~\ref{fig:cats in elliptic} summarizes the results for $\epsilon=\pi/5$ in the top row (a-c) and $\epsilon=\pi/7$ in the bottom row (d-f).

The first example $\epsilon=\pi/5$ is very close to the predicted boundary $\epsilon < \pi/4.78$ and the amplitude iso-surfaces in Fig.~\ref{fig:cats in elliptic}(a) do not visibly deviate from the two shapes familiar with radially-symmetric Gaussian beams; thus we might consider this case as of moderate deformation without complications. There are also no significant changes in the diagram in Fig.~\ref{fig:cats in elliptic}(b), which shows the number of rings in the parameter space $(P,\phi)$. But that is only because the change is too small to notice. In Fig.~\ref{fig:cats in elliptic}(c), we magnify a tiny region at the tip of the cusp of the gray line in Fig.~\ref{fig:cats in elliptic}(b), which is the bifurcation line of topological events on $y=0$, and see that it folds into a classic butterfly pattern. The range that the butterfly spans is so small that we could not have detected it, in both the $(P,\phi)$ parameter space and the amplitude iso-surfaces, without having prior knowledge about the catastrophes from Eq.~\eqref{eq:butterflyeqs}. The butterfly catastrophe changes the number of rings by 2, generating a small pair of rings off-axis that reconnect with the rest of the rings. 

\begin{figure*}
    \centering
    \includegraphics[width=0.99\textwidth]{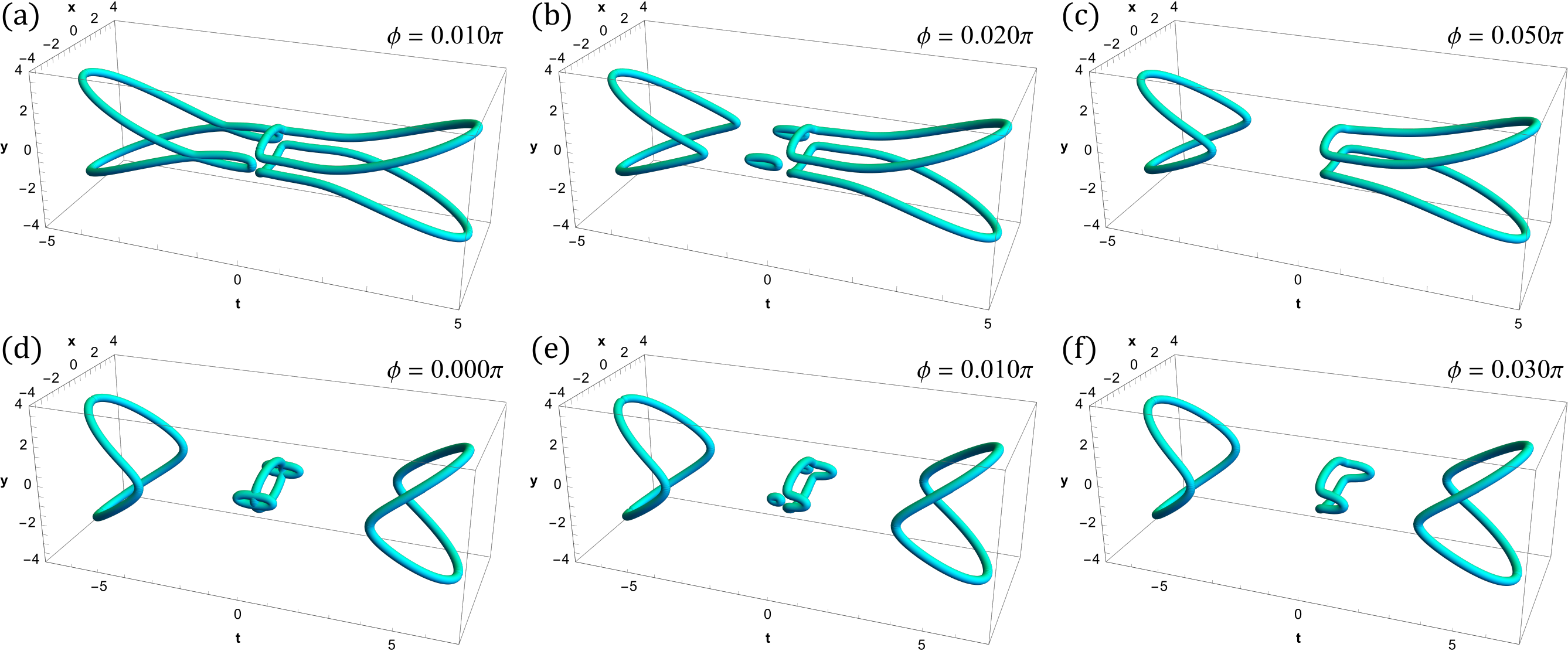}
    \caption{Reconnection of vortex rings in the elliptic beam with $d=0$ crossing the butterfly catastrophe at $\epsilon=\pi/7$ for (a-c) $P=1/7$ and (d-f) $P=1/9$.}
    \label{fig:elliptic rings}
\end{figure*}

Let us consider a more prominent example with $\epsilon=\pi/7$. The amplitude iso-surfaces in Fig.~\ref{fig:cats in elliptic}(d) now visibly deviate from the ones of moderate deformation - contours have a small ``bump'' at $t=0$. In Fig.~\ref{fig:cats in elliptic}(e), the folding of the bifurcation line of events on $y=0$ also spans a larger range of parameters. Again, in Fig.~\ref{fig:cats in elliptic}(f) we see that crossing the branches of the butterfly always changes the number of rings by 2.

Vortex rings transformation sequence depends on the path in parameter space, where we cross the butterfly. In Supplementary Video 1, we cross the ``wings'' of the butterfly, which in this case refers to the region with the two left-pointing cusps, $5.9<1/P<7.3$. Here, as we set $P=1/7$ and change $\phi$, the two small rings undergo $3\rightarrow5$ reconnection with a large single ring. The snapshots are presented in Fig.~\ref{fig:elliptic rings}(a-c), which shows two smaller rings breaking off the larger single ring until disappearing completely. 

In Supplementary Video 2, crossing the ``tail'' of the butterfly $7.3<1/P<14.3$ in Fig.~\ref{fig:cats in elliptic}(f) yields a slightly different result. The two small rings appear very close to the focal plane, and undergo $3\rightarrow1$ reconnection when they fuse with a smaller ring closer to $t=0$, as shown in the snapshots in Fig.~\ref{fig:elliptic rings}(d-f).

We demonstrated that a butterfly catastrophe appears for a very slight change of parameters but leads to a dramatic modification in vortex ring reconnections. Increasing $\epsilon$ further increases the span of parameters for the butterfly, but does not introduce any new bifurcations into the system beyond those already presented. Increasing the astigmatism parameter $d$, however, yields unexpected results, which are discussed in more detail in the next subsection.

\subsection{Astigmatic beams}

The astigmatism parameter separates the focal planes of $x$ and $y$ beam profiles by a distance $|d|$. Here, we focus on purely astigmatic beams ($\epsilon=\pi/4$), which have slightly different symmetries than the elliptic beam: the transverse profile is circular at $t=0$, and the semi-major axes of the transverse ellipses are orthogonal on the opposite sides of $t=0$.

Solving for Eqs.~\eqref{eq:cuspeqs} predicts that a pair of critical points should appear on-axis for $|d|>2$. In addition, both of Eqs.~\eqref{eq:butterflyeqs} predict the emergence of new critical points: a pair for $|d|>4.57$ and another pair for $|d|>4.76$, for each $f_x(t)$ in $y=0$ plane and $f_y(t)$ in $x=0$ plane. In total, therefore, we find 10 critical points by varying $d$. 

\begin{figure*}
    \centering
    \includegraphics[width=0.99\linewidth]{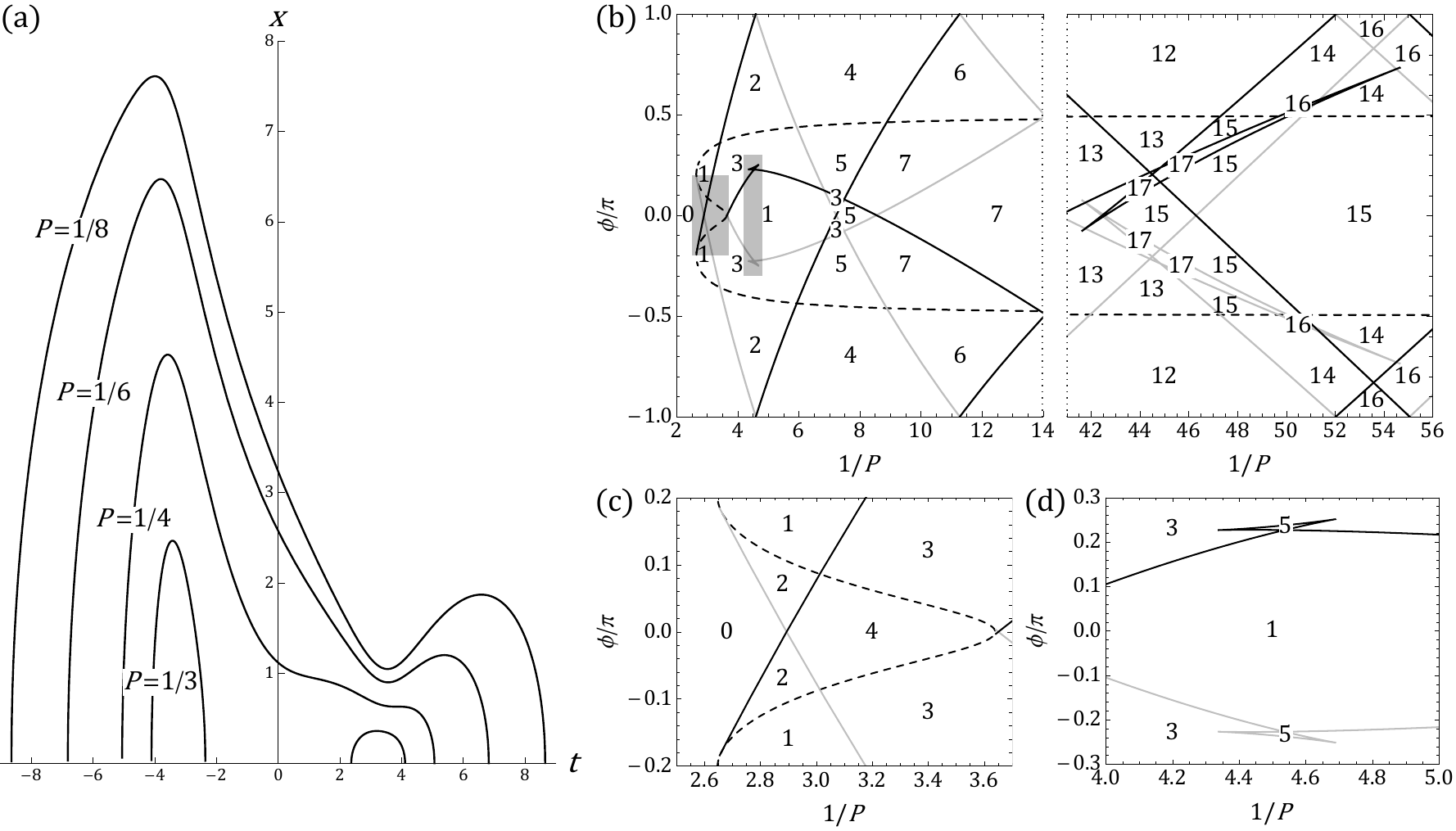}
    \caption{(a) Contours of amplitude iso-surfaces of the astigmatic beam with $\epsilon=\pi/4$ and $d=7$. (b) The number of vortex rings in the parameter domains $(P,\phi)$ separated by topological events located at the optical axis (dashed bifurcation lines), at the $x=0$ plane (black solid lines), and at the $y=0$ plane (grey solid lines). (c,d) Magnified grey regions of the parameter domain in (b).}
    \label{fig:cats in astigmatic}
\end{figure*}

Similar to the previous case of elliptic beams, we analyze the amplitude iso-surfaces, depicted in Fig.~\ref{fig:cats in astigmatic}(a) for $d=7$. Immediately we can see that the contours are deformed significantly. Notably, the iso-surface $P=1/3$ is split into two smaller spheroids, and $P=1/4$ has several points where the concavity flips, which appears to be similar to the ``bump'' in elliptic beams. These two ways of iso-surface deformation result in two different types of catastrophes. The latter one results in the butterfly catastrophe, as before, but the separation of amplitude iso-surfaces introduces a new type - the cusp catastrophe.

The cusp catastrophe starts at $|d|>2$, and, in the $(P,\phi)$ diagram in Fig.~\ref{fig:cats in astigmatic}(b), it appears as the inward folding of the dashed bifurcation line of events on the optical axis. Separation of amplitude iso-surfaces creates a region around $t=0$ where no rings can exist. Consequently, the cusp forces new topological transformations to occur at the inner endpoints of the iso-surfaces; mainly, there are two. In the domain $2<|d|<4$ these endpoints serve as nucleation points for a single ring. Supplementary Video 3 demonstrates an example of that in the astigmatic beam with $d=3$: a single ring disappears near $t=0$ and reappears on the other side. 

\begin{figure*}
    \centering
    \includegraphics[width=0.99\textwidth]{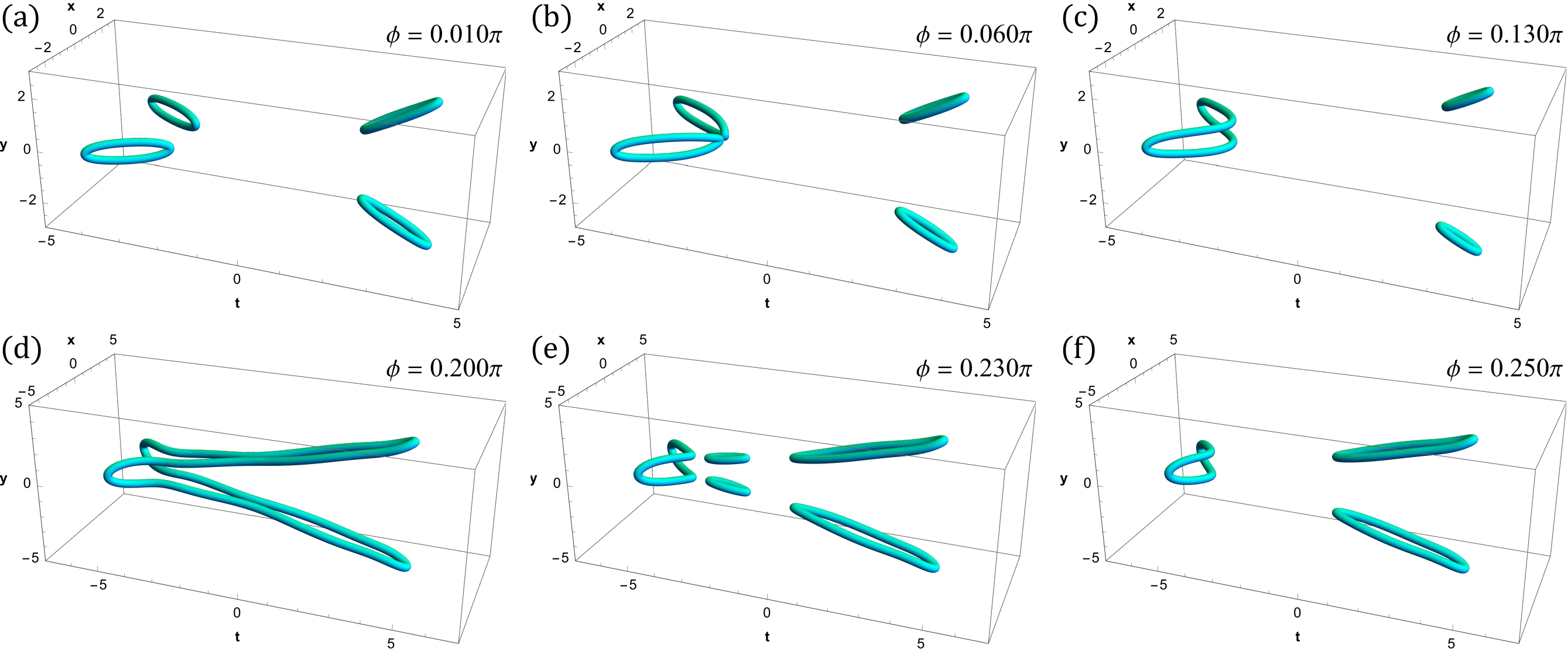}
    \caption{Reconnection of vortex rings in the astigmatic beam with the cusp and double butterfly catastrophe $\epsilon=\pi/4,d=7$: (a-c) $P=1/3.2$, (d-f) $P=1/4.5$.}
    \label{fig:astigmatic rings}
\end{figure*}

For sufficiently large astigmatism, such as $d=7$ in Fig.~\ref{fig:cats in astigmatic}(c), where the folding is magnified, we can see that crossing the dashed line changes the number of rings by 1, implying that reconnection can now occur on the optical axis. In Supplementary Video 4, we show that two rings fuse on a single point on the optical axis, as also shown in snapshots in Fig.~\ref{fig:astigmatic rings}(a-c). This type of reconnection - reconnection on-axis - is a completely new type of topological transformation that was not possible to achieve without the separation of amplitude surfaces for $|d|>4$.


The butterfly catastrophe in astigmatic beams manifests in a slightly different way - it decomposes into a swallowtail, but we still refer to it as ``the butterfly'' for consistency. The effects are the same: it generates a small pair of rings close to the focal planes. In fact, for $d=7$ there are two pairs of butterflies in bifurcation lines of reactions at $x=0$ and $y=0$ each in the $(P,\phi)$ space: the first smaller pair exist for $4.33<1/P<4.69$, depicted in Fig.~\ref{fig:cats in astigmatic}(d), and the other larger pair for $41.6<1/P<54.7$, shown in Fig.~\ref{fig:cats in astigmatic}(b). In both cases, crossing the butterfly changes the number of rings by 2.

However, in this case, choosing $P$ at which the bifurcation line intersects with itself enables a $1\rightarrow5$ reconnection, which is not possible in elliptic beams. An example of that is presented in Supplementary Video 5, with snapshots displayed in Fig.~\ref{fig:astigmatic rings}(d-f), where a large single ring breaks off into five, and the two smaller rings disappear quickly as $\phi$ changes.

After analyzing the effect of each beam parameter separately, we can finally construct the bigger picture - catastrophes in the full parameter domain $(\epsilon,d)$.

\subsection{Full parameter space}

\begin{figure}[h!]
    \centering
    \includegraphics[width=0.6\linewidth]{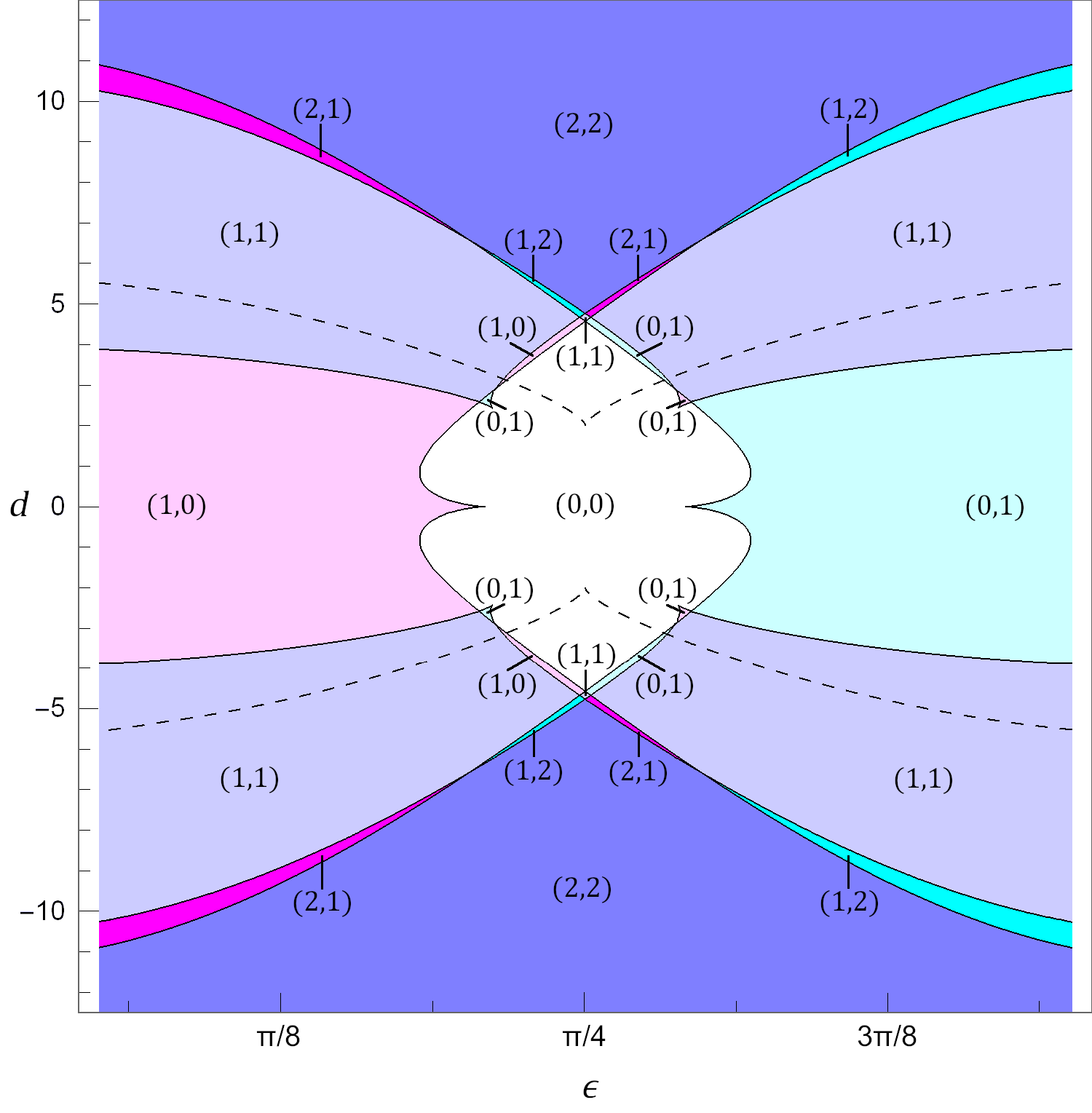}
    \caption{Domains with cusp and butterfly catastrophes in the parameter space $(\epsilon,d)$. The numbers in brackets indicate the numbers (0, 1, or 2) of butterfly catastrophes at, correspondingly, $(y=0, x=0)$ planes. Dashed lines separate the domains without cusp catastrophe (between the lines) from regions with the cusp (top and bottom).}
    \label{fig:catdiagram}
\end{figure}

We now solve for Eqs.~\eqref{eq:cuspeqs} and~\eqref{eq:butterflyeqs} without fixing $\epsilon$ or $d$ at any specific values. Each equation separately describes a three-dimensional surface $(t,\epsilon,d)$; the solution to the system is the intersection of two surfaces - lines in 3D. We obtain these lines numerically and project them onto a surface $(\epsilon,d)$, which produces a diagram presented in Fig. \ref{fig:catdiagram}. Here, the dashed lines are solutions to Eqs. \eqref{eq:cuspeqs}, separating domains with the cusp catastrophe, and the solid black - to Eqs. \eqref{eq:butterflyeqs}, which separate domains with the butterfly catastrophe.

The dashed lines of the cusp catastrophe are symmetric about $d=0$ and $\epsilon=\pi/4$, resembling the classic cusp pattern, hence the name. The lowest $|d|$ at which the catastrophe can occur is at $|d|>2$ for exactly $\epsilon=\pi/4$, which is a purely astigmatic beam. In combination with ellipticity, the value shifts up to $|d|>5.6$. As expected, the cusp catastrophe cannot be observed in purely elliptic beams $(d=0)$.

The solid lines separating domains with butterfly catastrophes appear a bit more complicated. The butterfly can appear once or twice on both $x=0$ and $y=0$, which corroborates our earlier findings. In purely elliptic beams, and even some elliptic-astigmatic beams, the butterfly catastrophe can happen only once on the $y=0$ plane if $\epsilon<\pi/4$ or the $x=0$ plane if $\epsilon>\pi/4$. In purely astigmatic beams, the butterflies always come in pairs: there is one pair of single butterflies for $4.57<|d|<4.76$, and a pair of double butterflies for $|d|>4.76$, as we have established earlier.

We emphasise that the domain free of any catastrophes, which is a white region in between the two dashed lines in Fig.~\ref{fig:catdiagram}, is small, especially for purely elliptic beams. This shows that increasing the beam parameters even slightly can trigger the folding of bifurcation lines of topological reactions, which in turn create new configurations of vortex rings that can be observed.

\section{Conclusions}

We investigate topological transformations of vortex rings in the superposition of the elliptical Gaussian beam and the plane wave. Beyond moderate deformation of the Gaussian beam, increasing the beam ellipticity $\epsilon$ and astigmatism $d$, we observe two types of catastrophes that affect the topological events: the cusp and the butterfly. 

The cusp catastrophe affects the reactions on the optical axis in two ways: by creating an additional pair of nucleation points or by allowing for on-axis reconnection. The cusp is the result of the separation of amplitude iso-surfaces, which is the direct consequence of the increased astigmatism $d$.

The butterfly catastrophe affects reactions off-axis and always generates an additional pair of small rings near the focal planes, although how they interact with the rest of the rings varies. This catastrophe is caused by the deformation of amplitude iso-surfaces around the focal planes, which causes the transverse intensity profile of the beam at the focal plane to appear as an elongated ellipse rather than a circular spot. 

We presented the study of an accessible system where complex topology of light can be realised in simple settings available in any optics laboratory with well established methods of detecting optical vortices and tracing their spatial configurations~\cite{FerrandoEtal2023}. Beyond optical vortices and structured light carrying singularities in the wavefronts of free-propagating laser beams, similar dislocations appear in the momentum space~\cite{Leykam} and geometric phase~\cite{Uakhitov,Zak} of optical fields propagating in structured materials, such as photonic lattices. Our results may be suggestive in designing photonic systems with phase dislocations control via nucleations and reconnections.

\section*{Conflict of Interest Statement}
The authors declare that the research was conducted in the absence of any commercial or financial relationships that could be construed as a potential conflict of interest.

\section*{Author Contributions}
ZK: Writing – original draft, Formal analysis, Investigation, Visualization; AF: Writing – review \& editing, Investigation; ASD: Writing – review \& editing, Conceptualization, Funding acquisition, Project administration, Supervision.

\section*{Funding}
ASD research is supported by Nazarbayev University FDCRGP grant 021220FD3551.


\section*{Supplemental Data}
 The Supplementary Videos for this article can be found online at: LINK by PUBLISHER

\bibliographystyle{Frontiers-Vancouver} 
\bibliography{cas-refs}

@Article{Adams2024,
  Title                    = {Spatiotemporal optical vortex reconnections of multi-vortices},
  Author                   = {Adams, J. and Agha, I. and Chong, A.},
  Journal                  = {Scientific Reports},
  Year                     = {2024},
  Volume                   = {14},
  Number                   = {1},
  Pages                    = {5483},
  Doi                      = {10.1038/s41598-024-54216-4},
  Numpages                 = {},
  Publisher                = {}
}

@article{AdamsChong2025,
    title = {Spatiotemporal optical vortex reconnections of loop vortices},
    author = {Adams, J. and Chong, A.},
    pages = {729--739},
    volume = {14},
    number = {6},
    journal = {Nanophotonics},
    doi = {10.1515/nanoph-2024-0594},
    year = {2025}
}

@Book{ArnoldWassermann2003,
  Title                    = {Catastrophe Theory, 3rd ed.},
  Author                   = {Arnold, V. and Wassermann, G.},
  Publisher                = {Springer},
  Year                     = {2003},
  Address                  = {},
  Doi                      = {10.1007/978-3-642-58124-3}
}

@Article{BazhenovVasnetsovSoskin1990,
  Title                    = {Laser beams with screw dislocations in their wavefronts},
  Author                   = {Bazhenov, V. and Vasnetsov, M. and Soskin, M.},
  Journal                  = {Optical and Quantum Electronics},
  Year                     = {1990},
  Volume                   = {24},
  Number                   = {},
  Pages                    = {951-962},
  Doi                      = {},
  Numpages                 = {},
  Publisher                = {}
}

@Article{Berry1976,
  Title                    = {Waves and {T}hom’s theorem},
  Author                   = {Berry, M. V.},
  Journal                  = {Advances in Physics},
  Year                     = {1976},
  Volume                   = {25},
  Number                   = {1},
  Pages                    = {1-26},
  Doi                      = {10.1080/00018737600101342},
  Numpages                 = {},
  Publisher                = {}
}

@Article{BerryUpstill1980,
  Title                    = {Catastrophe optics: morphologies of caustics and their diffraction patterns},
  Author                   = {Berry, M. V. and Upstill, C.},
  Journal                  = {Progress in Optics},
  Year                     = {1980},
  Volume                   = {28},
  Number                   = {},
  Pages                    = {257-346},
  Doi                      = {10.1016/S0079-6638(08)70215-4}
}

@Article{BerryNyeWright1978,
  Title                    = {The elliptic umbilic diffraction catastrophe},
  Author                   = {Berry, M. V. and Nye, J. and Wright, F.},
  Journal                  = {Philosophical Transactions of the Royal Society of London A},
  Year                     = {1978},
  Volume                   = {291},
  Number                   = {1382},
  Pages                    = {453-484},
  Doi                      = {10.1098/rsta.1979.0039},
  Numpages                 = {},
  Publisher                = {}
}

@Article{BerryDennis2001a,
  Title                    = {Knotted and linked phase singularities in monochromatic waves},
  Author                   = {Berry, M. V. and Dennis, M. R.},
  Journal                  = {Proceedings of the Royal Society of London A},
  Year                     = {2001},
  Volume                   = {457},
  Number                   = {},
  Pages                    = {2251-2263},
  Doi                      = {10.1098/rspa.2001.0826},
  Numpages                 = {},
  Publisher                = {}
}

@Article{BerryDennis2001b,
  Title                    = {Knotting and unknotting of phase singularities: Helmholtz waves, paraxial waves and waves in 2 + 1 spacetime},
  Author                   = {Berry, M. V. and Dennis, M. R.},
  Journal                  = {Journal of Physics A: Mathematical and general},
  Year                     = {2001},
  Volume                   = {34},
  Number                   = {},
  Pages                    = {8877-8888},
  Doi                      = {10.1088/0305-4470/34/42/311},
  Numpages                 = {},
  Publisher                = {}
}

@Article{BerryDennis2007,
  Title                    = {Topological events on wave dislocation lines: birth and death of loops, and reconnection},
  Author                   = {Berry, M. V. and Dennis, M. R.},
  Journal                  = {Journal of Physics A},
  Year                     = {2007},
  Volume                   = {40},
  Number                   = {},
  Pages                    = {65-74},
  Doi                      = {10.1088/1751-8113/40/1/004},
  Numpages                 = {},
  Publisher                = {}
}

@article{Bekshaev2024,
    author = {Bekshaev, A.},
    title = {Spatiotemporal optical vortices: Principles of description and basic properties},
    journal = {APL Photonics},
    volume = {9},
    number = {11},
    pages = {110806},
    year = {2024},
    month = {11},
    issn = {2378-0967},
    doi = {10.1063/5.0233758},
    url = {https://doi.org/10.1063/5.0233758},
    eprint = {https://pubs.aip.org/aip/app/article-pdf/doi/10.1063/5.0233758/20239331/110806_1_5.0233758.pdf},
}

@article{Bliokh2021,
  title = {Spatiotemporal Vortex Pulses: Angular Momenta and Spin-Orbit Interaction},
  author = {Bliokh, K. Y.},
  journal = {Phys. Rev. Lett.},
  volume = {126},
  issue = {24},
  pages = {243601},
  numpages = {6},
  year = {2021},
  month = {Jun},
  publisher = {American Physical Society},
  doi = {10.1103/PhysRevLett.126.243601},
  url = {https://link.aps.org/doi/10.1103/PhysRevLett.126.243601}
}

@Article{BoivinWolf1965,
  Title                    = {Electromagnetic field in the neighborhood of the focus of a coherent beam},
  Author                   = {Boivin, A. and Wolf, E.},
  Journal                  = {Physical Review},
  Year                     = {1965},
  Volume                   = {138},
  Number                   = {6B},
  Pages                    = {B1561–B1565},
  Doi                      = {10.1103/PhysRev.138.B1561},
  Numpages                 = {},
  Publisher                = {}
}

@Article{BoivinDowWolf1967,
  Title                    = {Energy Flow in the neighborhood of the focus of a coherent beam},
  Author                   = {Boivin, A. and Dow, J. and Wolf, E.},
  Journal                  = {Journal of the Optical Society of America},
  Year                     = {1967},
  Volume                   = {57},
  Number                   = {10},
  Pages                    = {1171–1175},
  Doi                      = {10.1364/JOSA.57.001171},
  Numpages                 = {},
  Publisher                = {}
}

@Article{Borghi2011,
  Title                    = {Evaluation of cuspoid and umbilic diffraction catastrophes of codimension four},
  Author                   = {Borghi, R.},
  Journal                  = {Journal of the Optical Society of America A},
  Year                     = {2011},
  Volume                   = {28},
  Number                   = {5},
  Pages                    = {887-896},
  Doi                      = {10.1364/JOSAA.28.000887},
  Numpages                 = {},
  Publisher                = {}
}

@Article{Borghi2016,
  Title                    = {Catastrophe optics of sharp-edge diffraction},
  Author                   = {Borghi, R.},
  Journal                  = {Optics Letters},
  Year                     = {2016},
  Volume                   = {41},
  Number                   = {13},
  Pages                    = {3114–3117},
  Doi                      = {10.1364/OL.41.003114},
  Numpages                 = {},
  Publisher                = {}
}

@Article{Borghi2022,
  Title                    = {Sharp-edge diffraction under Bessel beam illumination: a catastrophe optics perspective},
  Author                   = {Borghi, R. and Carosella, C.},
  Journal                  = {Journal of the Optical Society of America A},
  Year                     = {2022},
  Volume                   = {39},
  Number                   = {6},
  Pages                    = {1117–1127},
  Doi                      = {10.1364/JOSAA.455654},
  Numpages                 = {},
  Publisher                = {}
}

@article{ChenEtal2021,
    author={Chen, J. and Lu, K. and Cao, Q. and Wan, C. and Hu, H. and Zhan, Q.},
    title={Automated Close-Loop System for Three-Dimensional Characterization of Spatiotemporal Optical Vortex},
    journal={Frontiers in Physics},
    volume={9},
    year={2021},
    pages={633922},
    doi={10.3389/fphy.2021.633922},
    ISSN={2296-424X}
}

@Article{CoulletGilRocca2010,
title = {Optical vortices},
journal = {Optics Communications},
volume = {73},
number = {5},
pages = {403-408},
year = {1989},
issn = {0030-4018},
doi = {https://doi.org/10.1016/0030-4018(89)90180-6},
url = {https://www.sciencedirect.com/science/article/pii/0030401889901806},
author = {P. Coullet and L. Gil and F. Rocca}
}

@Article{DennisEtal2009,
  Title                    = {Singular Optics: more ado about nothing},
  Author                   = {Dennis, M. R. and Kivshar, Y. and Soskin, M. and Swartzlander Jr, G.},
  Journal                  = {Journal of Optics A: Pure and Applied Optics},
  Year                     = {2009},
  Volume                   = {11},
  Number                   = {9},
  Pages                    = {090201},
  Doi                      = {10.1088/1464-4258/11/9/090201},
  Numpages                 = {},
  Publisher                = {}
}

@Article{Desyatnikov2023,
  Title                    = {Vortex rings in paraxial laser beams},
  Author                   = {Desyatnikov, A. S.},
  Journal                  = {Optics Express},
  Year                     = {2023},
  Volume                   = {31},
  Number                   = {},
  Pages                    = {31955-31968},
  Doi                      = {10.1364/OE.500950}
}

@article{EilenbergerEtal2013,
  author  = {Eilenberger, F. and Prater, K. and Minardi, S. and Geiss, R. and R{\"o}pke, U. and Kobelke, J. and Schuster, K. and Bartelt, H. and Nolte, S. and T{\"u}nnermann, A. and Pertsch, T.},
  title   = {Observation of Discrete, Vortex Light Bullets},
  journal = {Physical Review X},
  volume  = {3},
  number  = {4},
  pages   = {041031},
  year    = {2013},
  doi     = {10.1103/PhysRevX.3.041031}
}

@Article{FerrandoEtal2023,
  Title                    = {Vortex-antivortex pair control in quadrupole {G}aussian beams},
  Author                   = {Ferrando, A. and Popiołek-Masajada, A. and Masajada, J. and Markevich, R. and Khoroshun, A.},
  Journal                  = {Optics Express},
  Year                     = {2023},
  Volume                   = {31},
  Number                   = {14},
  Pages                    = {23444},
  Doi                      = {10.1364/OE.494326}
}

@Article{ForbesOlivieraDennis2021,
  Title                    = {Structured light},
  Author                   = {Forbes, A. and de Oliviera, M. and Dennis, M. R.},
  Journal                  = {Nature Photonics},
  Year                     = {2021},
  Volume                   = {15},
  Number                   = {},
  Pages                    = {253-262},
  Doi                      = {10.1038/s41566-021-00780-4},
  Numpages                 = {},
  Publisher                = {}
}

@article{HancockEtal2024,
  title = {Spatiotemporal Torquing of Light},
  author = {Hancock, S. W. and Zahedpour, S. and Goffin, A. and Milchberg, H. M.},
  journal = {Phys. Rev. X},
  volume = {14},
  issue = {1},
  pages = {011031},
  numpages = {18},
  year = {2024},
  month = {Feb},
  publisher = {American Physical Society},
  doi = {10.1103/PhysRevX.14.011031},
  url = {https://link.aps.org/doi/10.1103/PhysRevX.14.011031}
}

@article{HancockZahedpourMilchberg2021,
  title = {Mode Structure and Orbital Angular Momentum of Spatiotemporal Optical Vortex Pulses},
  author = {Hancock, S. W. and Zahedpour, S. and Milchberg, H. M.},
  journal = {Phys. Rev. Lett.},
  volume = {127},
  issue = {19},
  pages = {193901},
  numpages = {6},
  year = {2021},
  month = {Nov},
  publisher = {American Physical Society},
  doi = {10.1103/PhysRevLett.127.193901},
  url = {https://link.aps.org/doi/10.1103/PhysRevLett.127.193901}
}

@Article{HöhmannEtal2010,
  Title                    = {Freak waves in the linear regime: a microwave study},
  Author                   = {Höhmann, R. and Kuhl, U. and Stöckmann, H.-J. and Kaplan, L. and Heller, E. J.},
  Journal                  = {Physical Review Letters},
  Year                     = {2010},
  Volume                   = {104},
  Number                   = {},
  Pages                    = {093901},
  Doi                      = {10.1103/PhysRevLett.104.093901},
  Numpages                 = {},
  Publisher                = {}
}

@article{JhajjEtal2016,
  author  = {Jhajj, N. and Larkin, I. and Rosenthal, E. W. and Zahedpour, S. and Wahlstrand, J. K. and Milchberg, H. M.},
  title   = {Spatiotemporal Optical Vortices},
  journal = {Physical Review X},
  volume  = {6},
  pages   = {031037},
  year    = {2016},
  doi     = {10.1103/PhysRevX.6.031037}
}

@Article{KulchukovaDesyatnikov2024,
  Title                    = {Vortex ring reconnections in elliptic Gaussian beams},
  Author                   = {Kulchukova, Z. and Desyatnikov, A. S.},
  Journal                  = {Optics Letters},
  Year                     = {2024},
  Volume                   = {49},
  Number                   = {4},
  Pages                    = {915-918},
  Doi                      = {10.1364/OL.515817},
  Numpages                 = {},
  Publisher                = {}
}

@article{LeEtal2024,
  title = {Self-Focused Pulse Propagation is Mediated by Spatiotemporal Optical Vortices},
  author = {Le, M. S. and Hine, G. A. and Goffin, A. and Palastro, J. P. and Milchberg, H. M.},
  journal = {Phys. Rev. Lett.},
  volume = {133},
  issue = {5},
  pages = {053803},
  numpages = {7},
  year = {2024},
  month = {Aug},
  publisher = {American Physical Society},
  doi = {10.1103/PhysRevLett.133.053803},
  url = {https://link.aps.org/doi/10.1103/PhysRevLett.133.053803}
}

@article{LiuCaoZhan2024,
    author = {Liu, X. and Cao, Q. and Zhan, Q.},
    title = {{Spatiotemporal optical wavepackets: from concepts to applications}},
    volume = {3},
    journal = {Photonics Insights},
    number = {4},
    publisher = {SPIE},
    pages = {R08},
    year = {2024},
    doi = {10.3788/PI.2024.R08},
    URL = {https://doi.org/10.3788/PI.2024.R08}
}

@Article{MathisEtal2015,
  Title                    = {Caustics and rogue waves in an optical sea},
  Author                   = {Mathis. A. and Froehly, L. and Toenger, S. and Dias, F. and Genty, G. and Dudley, J.},
  Journal                  = {Scientific Reports},
  Year                     = {2015},
  Volume                   = {5},
  Number                   = {},
  Pages                    = {12822},
  Doi                      = {10.1038/srep12822},
  Numpages                 = {},
  Publisher                = {}
}

@Book{Nye1999,
  Title                    = {Natural focusing and fine structure of light: caustics and wave dislocations},
  Author                   = {Nye, J.},
  Publisher                = {Institute of Physics Publishing},
  Year                     = {1999},
  Address                  = {Bristol},
  Doi                      = {}
}

@Article{Nye1998,
  Title                    = {Unfolding of higher-order wave dislocations},
  Author                   = {Nye. J.},
  Journal                  = {Journal of the Optical Society of America A},
  Year                     = {1998},
  Volume                   = {15},
  Number                   = {5},
  Pages                    = {1132-1138},
  Doi                      = {10.1364/JOSAA.15.001132},
  Numpages                 = {},
  Publisher                = {}
}

@Article{Nye2007,
  Title                    = {Dislocation lines in the swallowtail diffraction catastrophe},
  Author                   = {Nye, J.},
  Journal                  = {Proceedings of the Royal Society of London A},
  Year                     = {2007},
  Volume                   = {463},
  Number                   = {2078},
  Pages                    = {343–355},
  Doi                      = {10.1098/rspa.2006.1765},
  Numpages                 = {},
  Publisher                = {}
}

@Article{Nye2008,
  Title                    = {Unfolding higher-order wave dislocation clusters and catastrophe theory},
  Author                   = {Nye, J.},
  Journal                  = {Journal of Optics A: Pure and Applied Optics},
  Year                     = {2008},
  Volume                   = {10},
  Number                   = {7},
  Pages                    = {075010},
  Doi                      = {10.1088/1464-4258/10/7/075010},
  Numpages                 = {},
  Publisher                = {}
}

@Article{Nye1978,
  Title                    = {Optical caustics in the near field from liquid drops},
  Author                   = {Nye, J.},
  Journal                  = {Proceedings of the Royal Society of London A},
  Year                     = {1978},
  Volume                   = {361},
  Number                   = {1704},
  Pages                    = {21-41},
  Doi                      = {10.1098/rspa.1978.0090},
  Numpages                 = {},
  Publisher                = {}
}

@Article{Nye1986,
  Title                    = {The catastrophe optics of liquid drop lenses},
  Author                   = {Nye, J.},
  Journal                  = {Proceedings of the Royal Society of London A},
  Year                     = {1986},
  Volume                   = {1824},
  Number                   = {1704},
  Pages                    = {1-26},
  Doi                      = {10.1098/rspa.1986.0001},
  Numpages                 = {},
  Publisher                = {}
}

@Article{Nye2003,
  Title                    = {From {Airy} rings to the elliptic umbilic diffraction catastrophe},
  Author                   = {Nye, J.},
  Journal                  = {Journal of Optics A: Pure and Applied Optics},
  Year                     = {2003},
  Volume                   = {5},
  Number                   = {},
  Pages                    = {503-510},
  Doi                      = {10.1088/1464-4258/5/5/311},
  Numpages                 = {},
  Publisher                = {}
}

@Article{NyeBerry1974,
  Title                    = {Dislocations in wave trains},
  Author                   = {Nye, J. and Berry, M. V.},
  Journal                  = {Proceedings of the Royal Society of London A},
  Year                     = {1974},
  Volume                   = {336},
  Number                   = {},
  Pages                    = {165-190},
  Doi                      = {10.1098/rspa.1974.0012},
  Numpages                 = {},
  Publisher                = {}
}

@Article{Pearcey1946,
  Title                    = {{XXXI. The} structure of an electromagnetic field in the neighbourhood of a cusp of a caustic},
  Author                   = {Pearcey, T.},
  Journal                  = {The London, Edinburgh, and Dublin Philosophical Magazine and Journal of Science},
  Year                     = {1946},
  Volume                   = {37},
  Number                   = {268},
  Pages                    = {311-317},
  Doi                      = {10.1080/14786444608561335},
  Numpages                 = {},
  Publisher                = {}
}

@article{PetersCocotosForbes2025,
author = {Peters, C. and Cocotos, V. and Forbes, A.},
journal = {Adv. Opt. Photon.},
number = {1},
pages = {113--184},
publisher = {Optica Publishing Group},
title = {Structured light in atmospheric turbulence---a guide to its digital implementation: tutorial},
volume = {17},
month = {Mar},
year = {2025},
url = {https://opg.optica.org/aop/abstract.cfm?URI=aop-17-1-113},
doi = {10.1364/AOP.538883},
}

@article{Porras2024,
doi = {10.1088/2040-8986/ad645a},
url = {https://doi.org/10.1088/2040-8986/ad645a},
year = {2024},
month = {jul},
publisher = {IOP Publishing},
volume = {26},
number = {9},
pages = {095601},
author = {Porras, M. A.},
title = {Clarification of the transverse orbital angular momentum of spatiotemporal optical vortices},
journal = {Journal of Optics}
}

@Article{Rubinsztein-DunlopEtal2017,
  Title                    = {Roadmap on structured light},
  Author                   = {Rubinsztein-Dunlop, H. and Forbes, A. and Berry, M. V. and Dennis, M. R. and Andrews, D. L. and Mansuripur, M. and Denz, C. and Alpmann, C. and Banzer, P. and Bauer, T. and Karimi, E. and Marrucci, L. and Padgett, M. and Ritsch-Marte, M. and Litchinitser, N. M. and Bigelow, N. P. and Rosales-Guzmán, C. and Belmonte, A. and Torres, J. P. and Neely, T. W. and Baker, M. and Gordon, R. and Stilgoe, A. B. and Romero, J. and White, A. G. and Fickler, R. and Willner, A. E. and Xie, G. and McMorran, B. and Weiner, A. M.},
  Journal                  = {Journal of Optics},
  Year                     = {2017},
  Volume                   = {19},
  Number                   = {},
  Pages                    = {013001},
  Doi                      = {10.1088/2040-8978/19/1/013001},
  Numpages                 = {},
  Publisher                = {}
}

@article{ShenEtal2023,
    doi = {10.1088/2040-8986/ace4dc},
    url = {https://doi.org/10.1088/2040-8986/ace4dc},
    year = {2023},
    month = {},
    publisher = {IOP Publishing},
    volume = {25},
    number = {9},
    pages = {093001},
    author = {Shen, Y. and Zhan, Q. and Wright, L. G. and Christodoulides, D. N. and Wise, F. W. and Willner, A. E. and Zou, K. and Zhao, Z. and Porras, M. A. and Chong, A. and Wan, C. and Bliokh, K. Y. and Liao, C. and Hernández-García, C. and Murnane, M. and Yessenov, M. and Abouraddy, A. F. and Wong, L. J. and Go, M. and Kumar, S. and Guo, C. and Fan, S. and Papasimakis, N. and Zheludev, N. I. and Chen, L. and Zhu, W. and Agrawal, A. and Mounaix, M. and Fontaine, N. K. and Carpenter, J. and Jolly, S. W. and Dorrer, C. and Alonso, B. and Lopez-Quintas, I. and López-Ripa, M. and Sola, Í. J. and Huang, J. and Zhang, H. and Ruan, Z. and Dorrah, A. H. and Capasso, F. and Forbes, A.},
    title = {Roadmap on spatiotemporal light fields},
    journal = {Journal of Optics},
}

@Article{ShenEtal2019,
  Title                    = {Optical vortices 30 years on: {OAM} manipulation from topological charge to multiple singularities},
  Author                   = {Shen, Y. and Wang, X. and Xie, Z. and Min, C. and Fu, X. and Liu, Q. and Gong, M. and Yuan, X.},
  Journal                  = {Light: Science \& Applications},
  Year                     = {2019},
  Volume                   = {8},
  Number                   = {90},
  Pages                    = {},
  Doi                      = {10.1038/s41377-019-0194-2},
  Numpages                 = {},
  Publisher                = {}
}

@Article{SoskinEtal2017,
  Title                    = {Singular optics and topological photonics},
  Author                   = {Soskin, M. and Boriskina, S. and Chong, Y. and Dennis, M. R. and Desyatnikov, A. S.},
  Journal                  = {Journal of Optics},
  Year                     = {2017},
  Volume                   = {19},
  Number                   = {},
  Pages                    = {010401},
  Doi                      = {10.1088/2040-8986/19/1/010401},
  Numpages                 = {},
  Publisher                = {}
}

@Article{TrinkausDrepper1977,
  Title                    = {On the analysis of diffraction catastrophes},
  Author                   = {Trinkaus, H. and Drepper, F.},
  Journal                  = {Journal of Physics A Mathematical and General},
  Year                     = {1977},
  Volume                   = {10},
  Number                   = {1},
  Pages                    = {L11–L16},
  Doi                      = {10.1088/0305-4470/10/1/003},
  Numpages                 = {},
  Publisher                = {}
}

@article{Tripathi2025,
author = {Tripathi, N. and Hancock, S. W. and Milchberg, H. M.},
journal = {Opt. Express},
number = {20},
pages = {41766--41776},
publisher = {Optica Publishing Group},
title = {Transverse orbital angular momentum of spatiotemporal optical vortices: setting the record straight},
volume = {33},
month = {Oct},
year = {2025},
url = {https://opg.optica.org/oe/abstract.cfm?URI=oe-33-20-41766},
doi = {10.1364/OE.568833},
}

@Book{Thom1972,
  Title                    = {Structural Stability and Morphogenesis},
  Editor                   = {},
  Author                   = {Thom, R.},
  Publisher                = {W. A. Benjamin},
  Year                     = {1972},
  Address                  = {},
  Doi                      = {10.1201/9780429493027}
}

@Article{TengEtal2021,
  Title                    = {Swallowtail-type diffraction catastrophe beams},
  Author                   = {Teng, H. and Qian, Y. and Lan, Y. and Cui, W.},
  Journal                  = {Optics Express},
  Year                     = {2021},
  Volume                   = {29},
  Number                   = {3},
  Pages                    = {3786–3794},
  Doi                      = {10.1364/OE.416134},
  Numpages                 = {},
  Publisher                = {}
}

@article{VoBliokhAlonso2026,
  title = {Orbital angular momentum of spatiotemporal vortices: A ray-mechanical analogy},
  author = {Vo, S. and Bliokh, K. Y. and Alonso, M. A.},
  journal = {Phys. Rev. A},
  volume = {114},
  issue = {1},
  pages = {013508},
  numpages = {10},
  year = {2026},
  month = {Jul},
  publisher = {American Physical Society},
  doi = {10.1103/hvqp-p3kl},
  url = {https://link.aps.org/doi/10.1103/hvqp-p3kl}
}

@incollection{Wolf2001,
  author    = {Soskin, M. and Vasnetsov, M.},
  title     = {{Chapter 4 - Singular Optics}},
  booktitle = {Progress in Optics},
  editor    = {Wolf, Emil},
  volume    = {42},
  pages     = {219-276},
  year      = {2001},
  publisher = {Elsevier}
}

@incollection{PiO2005,
  author    = {Desyatnikov, Anton S. and Kivshar, Yuri S. and Torner, Lluis},
  title     = {{Chapter 5 - Optical Vortices and Vortex Solitons}},
  booktitle = {Progress in Optics},
  editor    = {Wolf, Emil},
  volume    = {47},
  pages     = {291--391},
  year      = {2005},
  publisher = {Elsevier},
  issn      = {0079-6638},
  doi       = {10.1016/S0079-6638(05)47006-7}
}

@Article{Wan2022,
  Title                    = {Toroidal vortices of light},
  Author                   = {Wan, C. and Cao, Q. and Chen, J. and Chong, A. and Zhan, Q.},
  Journal                  = {Nature Photonics},
  Year                     = {2022},
  Volume                   = {16},
  Number                   = {},
  Pages                    = {519–522},
  Doi                      = {10.1038/s41566-022-01013-y},
  Numpages                 = {},
  Publisher                = {}
}

@Article{ZannottiEtal2017,
  Title                    = {Optical catastrophes of the swallowtail and butterfly beams},
  Author                   = {Zannotti, A. and Diebel, F. and Boguslawski, M. and Denz, C.},
  Journal                  = {New Journal of Physics},
  Year                     = {2017},
  Volume                   = {19},
  Number                   = {},
  Pages                    = {053004},
  Doi                      = {10.1088/1367-2630/aa6ecd},
  Numpages                 = {},
  Publisher                = {}
}

@Article{ZannottiDiebelDenz2017,
  Title                    = {Dynamics of the optical swallowtail catastrophe},
  Author                   = {Zannotti, A. and Diebel, F. and Denz, C.},
  Journal                  = {Optica},
  Year                     = {2017},
  Volume                   = {4},
  Number                   = {10},
  Pages                    = {1157–1162},
  Doi                      = {10.1364/OPTICA.4.001157},
  Numpages                 = {},
  Publisher                = {}
}

@Article{ZannottiEtal2020,
  Title                    = {Shaping caustics into propagation-invariant light},
  Author                   = {Zannotti, A. and Denz, C. and Alonso, M. and Dennis, M. R.},
  Journal                  = {Nature Communications},
  Year                     = {2020},
  Volume                   = {11},
  Number                   = {},
  Pages                    = {3597},
  Doi                      = {10.1038/s41467-020-17439-3},
  Numpages                 = {},
  Publisher                = {}
}

@incollection{Zeeman1979,
  author    = {Zeeman, E. C.},
  title     = {Catastrophe theory},
  booktitle = {Structural Stability in Physics},
  editor    = {G{\"u}ttinger, Werner and Eikemeier, Horst},
  publisher = {Springer},
  year      = {1979},
  pages     = {12--22}
}

@inproceedings{Zham25,
  author    = {Kulchukova, Zhamila and Ferrando, Albert and Desyatnikov, Anton},
  title     = {Optical Vortex Rings: Reconnections and Catastrophes},
  booktitle = {Proceedings of the 2025 Conference on Lasers and Electro-Optics Europe \& European Quantum Electronics Conference (CLEO/Europe-EQEC)}, 
  year      = {2025},
  pages     = {1},
  publisher = {IEEE}
}

@inproceedings{Zham26,
  author       = {Desyatnikov, Anton and Kulchukova, Zhamila and Ferrando, Albert},
  title        = {Accessible vortex rings in paraxial optics},
  booktitle    = {Complex Light and Optical Forces XX},
  editor       = {Andrews, David L. and Galvez, Enrique J. and Rubinsztein-Dunlop, Halina},
  volume       = {PC13921},
  pages        = {{PC139210J}},
  year         = {2026},
  organization = {International Society for Optics and Photonics},
  publisher    = {SPIE},
  doi          = {10.1117/12.3080459}
}

@Article{Zak,
AUTHOR = {Uakhitov, Tileubek and Urmanov, Abdybek and Kumekov, Serik E. and Desyatnikov, Anton S.},
TITLE = {Zak-Phase Dislocations in Trimer Lattices},
JOURNAL = {Symmetry},
VOLUME = {17},
YEAR = {2025},
NUMBER = {10},
ARTICLE-NUMBER = {1631},
URL = {https://www.mdpi.com/2073-8994/17/10/1631},
ISSN = {2073-8994},
DOI = {10.3390/sym17101631}
}

@article{Leykam,
author = {Daniel Leykam and Anton S. Desyatnikov},
title = {Conical intersections for light and matter waves},
journal = {Advances in Physics: X},
volume = {1},
number = {1},
pages = {101--113},
year = {2016},
publisher = {Taylor \& Francis},
doi = {10.1080/23746149.2016.1144482}
}

@inproceedings{Uakhitov,
  author    = {Uakhitov, Tileubek and Urmanov, Abdybek and Kumekov, Serik E. and Desyatnikov, Anton S.},
  title     = {Geometric Phase Dislocations in One-Dimensional Lattices},
  booktitle = {Proceedings of the 12th International Conference on Metamaterials, Photonic Crystals and Plasmonics (META)},
  year      = {2022},
  pages     = {218--219},
  url       = {https://www.scopus.com/pages/publications/85179875404?origin=resultslist}
}

@article{Knots2012,
  author = {Desyatnikov, Anton S. and Buccoliero, Daniel and Dennis, Mark R. and Kivshar, Yuri S.},
  title = {Spontaneous knotting of self-trapped waves},
  journal = {Scientific Reports},
  year = {2012},
  volume = {2},
  number = {1},
  pages = {771},
  doi = {10.1038/srep00771},
  url = {https://doi.org/10.1038/srep00771},
  isbn = {2045-2322}
}

@article{PRE2000,
  title = {Three-dimensional spinning solitons in dispersive media with the cubic-quintic nonlinearity},
  author = {Desyatnikov, Anton and Maimistov, Andrey and Malomed, Boris},
  journal = {Physical Review E},
  volume = {61},
  issue = {3},
  pages = {3107--3113},
  year = {2000},
  publisher = {American Physical Society},
  doi = {10.1103/PhysRevE.61.3107},
  url = {https://link.aps.org/doi/10.1103/PhysRevE.61.3107}
}

@article{Malomed2005,
doi = {10.1088/1464-4266/7/5/R02},
url = {https://doi.org/10.1088/1464-4266/7/5/R02},
year = {2005},
volume = {7},
number = {5},
pages = {R53},
author = {Malomed, Boris A and Mihalache, Dumitru and Wise, Frank and Torner, Lluis},
title = {Spatiotemporal optical solitons},
journal = {Journal of Optics B: Quantum and Semiclassical Optics}
}

@article{MalomedPRE2000,
  title = {{Stable vortex solitons in the two-dimensional Ginzburg-Landau equation}},
  author = {Crasovan, L.-C. and Malomed, B. A. and Mihalache, D.},
  journal = {Physical Review E},
  volume = {63},
  issue = {1},
  pages = {016605},
  numpages = {6},
  year = {2000},
  month = {Dec},
  publisher = {American Physical Society},
  doi = {10.1103/PhysRevE.63.016605},
  url = {https://link.aps.org/doi/10.1103/PhysRevE.63.016605}
}

\end{document}